\documentclass[aps,pra,twocolumn,byrevtex,showpacs,floatfix,superscriptaddress,reprint]{revtex4-2}
\usepackage{graphicx}
\usepackage{subcaption}
\usepackage{amsmath,amsfonts,amsthm,bm}
\usepackage{epsfig}
\usepackage{hyperref}

\usepackage{caption,subcaption}
\usepackage{mathtools}

\newcommand{\ket}[1]{\ensuremath{|\,{#1}\,\rangle}}

\hypersetup{colorlinks=true,linkcolor=blue,citecolor=blue,urlcolor=blue}

\begin{document}

\title{Training a quantum measurement device to discriminate unknown non-orthogonal quantum states}

\author{D.~Concha}
\email[corresponding author: ]{dconcham@udec.cl}
\affiliation{Instituto Milenio de Investigaci\'on en \'Optica y Departamento de F\'isica, Universidad de Concepci\'on, casilla 160-C, Concepci\'on, Chile}

\author{L. Pereira}
\affiliation{Instituto de F\'{\i}sica Fundamental IFF-CSIC, Calle Serrano 113b, Madrid 28006, Spain}

\author{L.~Zambrano}
\affiliation{Instituto Milenio de Investigaci\'on en \'Optica y Departamento de F\'isica, Universidad de Concepci\'on, casilla 160-C, Concepci\'on, Chile}

\author{A.~Delgado}
\affiliation{Instituto Milenio de Investigaci\'on en \'Optica y Departamento de F\'isica, Universidad de Concepci\'on, casilla 160-C, Concepci\'on, Chile}

\date{\today}


\begin{abstract}

Here, we study the problem of decoding information transmitted through unknown quantum states. We assume that Alice encodes an alphabet into a set of orthogonal quantum states, which are then transmitted to Bob. However, the quantum channel that mediates the transmission maps the orthogonal states into non-orthogonal states, possibly mixed. If an accurate model of the channel is unavailable, then the states received by Bob are unknown. In order to decode the transmitted information we propose to train a measurement device to achieve the smallest possible error in the discrimination process. This is achieved by supplementing the quantum channel with a classical one, which allows the transmission of information required for the training, and resorting to a noise-tolerant optimization algorithm. We demonstrate the training method in the case of minimum-error discrimination and show that it achieves error probabilities very close to the optimal one. In particular, in the case of two unknown pure states our proposal approaches the Helstrom bound. A similar result holds for a larger number of states in higher dimensions. We also show that a reduction of the search space, which is used in the training process, leads to a considerable reduction in the required resources. Finally, we apply our proposal to the case of the phase flip channel reaching an accurate value of the optimal error probability.

\end{abstract}

\maketitle

\section{Introduction}

The states of quantum systems have properties that distinguish them from their classical counterparts. Unknown quantum states cannot be perfectly and deterministically copied \cite{Wootters} and entangled states exhibit correlations without classical equivalence \cite{Bell}. These are deeply related to the impossibility of discriminating non-orthogonal quantum states. If this were the case, then unknown quantum states could be perfectly and deterministically copied and entangled states could be used to implement super luminal communications \cite{Gisin}. Consequently, the discrimination of non-orthogonal quantum states has become an important research subject due to its implications for the foundations of the quantum theory \cite{Pusey,Schmid} and quantum communications \cite{Chefles1,Bae}. For example, the problem of implementing quantum teleportation, entanglement sharing, and dense coding through a partially entangled pure state can be solved by local discrimination of non-orthogonal pure states \cite{PhysRevA.71.012303,PhysRevA.89.012337,PhysRevA.85.062322,PhysRevA.68.022310,PhysLettA.239,PhysRevA.87.052327,Solis,Pati}.

The discrimination of quantum states can be naturally stated in the context of two parties that attempt to communicate: Alice encodes information representing the letters of an alphabet through a set $\Omega_1$ of orthogonal pure quantum states, which is transmitted through a communication channel. The channel transforms the orthogonal states into a new set $\Omega_2$ of states, which might become non-orthogonal and mixed. Bob then receives them and performs a generalized quantum measurement to discriminate the states and to retrieve the information encoded by Alice. Both parties are assumed to know the generation probabilities of the orthogonal states in $\Omega_1$ and the set $\Omega_2$ of non-orthogonal states in advance. To decode the transmitted information, the parties agree on a figure of merit which is subsequently optimized to obtain the best single-shot generalized quantum measurement. This leads to several discrimination strategies such as minimum-error discrimination \cite{Holevo,Yuen,Helstrom}, pretty-good measurement \cite{PGM1,PGM2}, unambiguous discrimination \cite{Ivanovic,Dieks,Peres,Chefles2}, maximum-confidence discrimination \cite{Croke,Jimenez}, and fixed-rate of inconclusive result \cite{Bagan,Herzog}. Various discrimination strategies have already been experimentally demonstrated \cite{Cook,Barnett,Clarke,Mohseni,Waldherr,Prosser,Sol_s_Prosser_2021,PhysRevA.94.042309}. 

Typical figures of merit for state discrimination are functions of the generation probabilities and the conditional probabilities between the states transmitted by Alice and Bob's measurement outcomes, which in turn also depend on the generalized measurement used by Bob and the states received by Bob. Thereby, optimizing any figure of merit and choosing the best generalized measurement become difficult problems. This imposes the use of numerical optimization techniques such as semidefinite programming (SDP)\cite{holevo1973probl,eldar2003designing,eldar2003semidefinite}. 

Recently, the distinguishability between two known families of non-orthogonal quantum states has been studied by training quantum circuits through neural networks \cite{HChen,Patterson}. Quantum state discrimination has also been studied in the context of unknown quantum states. In this case, the communication channel maps the states in $\Omega_1$ into a new set of states $\Omega_2$ that are unknown to the communicating parties. An example of this situation is free-space quantum communication \cite{Ursin,Toyoshima,Jin,Steinlechner,Liao}, where information is encoded into states of light and transmitted through the atmosphere. This exhibits local and temporal variations in the refractive index, which can greatly modify the state of light and makes it difficult to characterize the transmitted states. Given that neither Alice nor Bob have access to the density matrices of the communicated states, standard approaches can not be applied. This problem has been studied assuming that the unknown states can be stored in quantum registers that control the action of a measurement device \cite{Dusek,Bergou1,Hayashi,Bergou2,Probst-Schendzielorz,Zhou}, or programmable discriminator.

In this article, we propose the training of a measurement device to optimally discriminate a set of unknown non-orthogonal quantum states. We assume that the action of this device is defined by a large set of control parameters, such that a given set of parameter values corresponds to the realization of a positive operator-valued measure (POVM). Given a fixed figure of merit for the discrimination process, it is iteratively optimized in the space of the control parameters. The optimization is driven by a gradient-free stochastic optimization algorithm \cite{Utreras,Zambrano,https://doi.org/10.48550/arxiv.2203.06044}, which approximates the gradient of the figure of merit by a finite difference. This requires at each iteration evaluations of the figure of merit at two different points in the control parameter space. Thereby, the training is driven by experimentally acquired data. Furthermore, stochastic optimization methods have been shown to be robust against noise \cite{Rambach2021}, so they are a standard choice in experimental contexts. The training of the measurement device is carried out until approaching the optimal value of the figure of merit within a prescribed tolerance. 

We illustrate our approach by studying minimum-error discrimination, where the figure of merit is the average retrodiction error. This figure of merit can be experimentally evaluated if, during the training step, Alice communicates the labels of the states that she sent to Bob through a classical channel. Minimum-error discrimination plays key role in quantum imaging \cite{Tan}, quantum reading \cite{Pirandola}, image discrimination \cite{Nair}, error-correcting codes \cite{Lloyd}, and quantum repeaters \cite{Loock}. This problem does not have a closed analytical solution except for sets of states with high symmetry. Our approach may also implement other discrimination strategies at the expense of resorting to more elaborate optimization algorithms. We first consider the minimum-error discrimination of two unknown non-orthogonal pure states. In this case the minimum of the average error probability, which can be analytically calculated, is given by the Helstrom bound. We show that it is possible to train the measurement device to reach values very close to the Helstrom bound. We extend our analysis to $d$ unknown non-orthogonal quantum states using $d$-dimensional symmetric states. Discrimination of this class of states plays an important role in processes such as quantum teleportation \cite{PhysRevA.68.022310,PhysRevA.85.062322}, entanglement swapping \cite{Solis}, and dense coding \cite{Pati} when carried out with partially entangled states and has already been implemented experimentally \cite{Prosser}. In this case, our approach also leads to the optimal single-shot generalized quantum measurement. However, it requires a large number of iterations. This is a consequence of the dimension of the control parameter space that scales as $d^4$. We also show that the use of a priori information effectively reduces the number of iterations, where we consider the use of initial conditions close to the optimal measurement as well as the reduction of the dimension of the control parameter space by assuming a particular property of the optimal measurement. Finally, we consider the discrimination of unknown mixed quantum states generated by quantum channels such as phase flip. 

This article is organized as follows: in Sec. II we introduce our approach to the discrimination of unknown orthogonal quantum states. In Sec. III we study the properties of our approach by means of several numerical experiments. In Sec. IV we summarize and conclude.

\section{Method}

Alice encodes the information to be transmitted into a set $\Omega_1=\{|\psi_q\rangle\}$ of $N$ mutually orthogonal pure states that are generated with probabilities $\{\eta_q\}$. The communication channel transforms the states in $\Omega_1$ into the states $\{\rho_q\}$ in $\Omega_2$, pure or mixed. For simplicity, we assume that the relation between states in $\Omega_1$ and $\Omega_2$ is one to one and that the action of the channel does not change the generation (or a priori) probabilities. Upon receiving each state, Bob tries to decode the information sent by Alice using a positive operator-valued measure $\{E_m\}$ composed of positive semi-definite matrices $E_m$ such that $\sum_mE_m=\emph{I}$, the identity operator. The probability of obtaining the $m$-th measurement outcome given that the state $\rho_q$ was sent is $P(E_m|\rho_q)=Tr(E_m\rho_q)$. We assume that if Bob obtains the $m$-th measurement result, he concludes that Alice attempted to transmit the state $|\psi_m\rangle$. This decoding rule leads to errors unless the states in $\Omega_2$ are mutually orthogonal, which leads Bob to seek to minimize the occurrence of errors in the discrimination process. Thereby, Bob needs to find the optimal POVM that minimizes the figure of merit that accounts for the errors.

Several quantum state discrimination strategies are known, each defined by a particular figure of merit. Here, we focus on minimum-error discrimination, where the number of states to be identified is equal to the number of elements of the POVM. The probability of correctly identifying the state $|\psi_q\rangle$ is given by $P(E_q|\rho_q)$. Since the states in $\Omega_2$ are generated with probabilities $\{\eta_q\}$, the average probability of correctly identifying all states is given by 
\begin{equation}
p_{corr}=\sum_{l=1}^N\eta_lTr(E_l\rho_l).
\end{equation}
The average error probability is $p_{err}=1-p_{corr}$. This probability, which is a function of the POVM $\{E_m\}$ and of the unknown states $\{\rho_k\}$, is minimized over the POVM space in order to train the measurement device. We assume that the states are fixed, that is, every time Alice aims to transmit the state $|\psi_k\rangle$, Bob receives the same state $\rho_k$. Thereby, the set $\Omega_2$ behaves as a set of unknown fixed parameters. 

The value of $p_{err}$ corresponds to a sum over probabilities $\eta_lTr(\rho_lE_l)$, which must be independently estimated. In order to do this Alice sends $N$ copies of each state $|\psi_l\rangle$ to Bob, communicating classically the label of them. These states play the role of training set. Bob measures the states with the corresponding POVM, which allows estimate the value of $Tr(\rho_lE_l)$. We simulate the experiment that allows us to estimate this value. For simulation purposes the states $\{\rho_l\}$ are known and thus we calculate the probabilities
\begin{equation}
p_l=Tr(\rho_lE_l)~{\rm and}~q_l=1-p_l=1-\sum_{m\neq l}Tr(\rho_kE_m).
\end{equation}
These are employed to generate a random number $n_l$ from a binomial distribution with success probability $p_l$ on a sample of size $N$. The probability $p_l$ is estimated as $n_l/N$. This procedure is repeated for each state in the set $\{\rho_k\}$ and for each one of the POVMs at each iteration. Thereby, the average error probability is estimated as
\begin{equation}
p_{err}=1-\sum_l\eta_l\frac{n_l}{N}.
\label{Estimate-minimum-error}
\end{equation}

We consider that the measurement device implements a POVM using the direct sum extension. According to this, the Hilbert space ${\cal H}_s$ of the states to be discriminated is complemented with an ancilla space ${\cal H}_a$, obtaining an extended Hilbert space ${\cal H}_e={\cal H}_s\oplus{\cal H}_a$. The POVM is implemented by applying a unitary transformation $U$ on ${\cal H}_e$ followed by a projective measurement on ${\cal H}_e$. This procedure requires adding fewer dimensions than the extension by means of the tensor product \cite{Chen}. To generate the unitary matrix $U$ we use a complex matrix $Z$ of order $dn\times d$, where $d$ is the dimension of ${\cal H}$ and $n$ is the number of states to be discriminated, and through the QR decomposition, we project it into an isometric matrix $S$, that is, a matrix such $S^\dagger S = I_{d\times d}$. Reshaping $S$ into a rank-3 tensor $S_{ijk}$ of size $n\times d\times d$, we can generate a full-rank POVM as $E_i = M_i^\dag M_i$, where the components of the matrices $M_i$ are $M_{i,jk}=S_{ijk}$. To physically realize the unitary transformation $U$ of order $dn\times dn$ one would need to fill the matrix $S$ with free parameters determined only up to unitarity. The average error probability $p_{err}$ can be thus regarded as a function $f(\bm z)$ of the complex vector ${\bm z}$ whose coefficients are given by the matrix elements of $U$, that is, we have $f(\bm z)=p_{err}({\bm z})$. 

We assume that neither Alice nor Bob knows the states in $\Omega_2$. Therefore, we cannot numerically evaluate the error probability $p_{err}$ or its derivatives. Besides, given that we consider POVMs implemented by direct sum extension, the shift parameter rule can not be applied to evaluate the gradient \cite{Li2017, Schuld2019}. Therefore, the optimization problem cannot be carried out using SDP or gradient-based optimization algorithms. To overcome this problem, we optimize $p_{err}$ using the Complex simultaneous perturbation stochastic approximation (CSPSA) \cite{Utreras,Zambrano,https://doi.org/10.48550/arxiv.2203.06044}, a gradient-free algorithm. This is based on the iterative rule
\begin{equation}
{\bm z}_{k+1}={\bm z}_{k}-a_k{\bm g}_{k}
\label{RULE}
\end{equation}
where ${\bm z}_k$ is a complex vector in the control parameter space at the $k$-th iteration, $a_k$ is a positive gain coefficient, and ${\bm g}_{k}$ is an approximation of the gradient of the figure of merit $f({\bm z})$  whose components are given by
\begin{equation}
g_{k,i}=\frac{f({\bm z_{k,+}})-f({\bm z_{k,-}})+\zeta_{k,+}-\zeta_{k,+}}{2c_k\Delta^*_{k,i}}.
\label{GRADIENT}
\end{equation}
In the expression above, the quantities $f({\bm z_{k,+}})$ and $f({\bm z_{k,-}})$ are the values of the figure of merit on the vectors
\begin{equation}
{\bm z_{k,\pm}}={\bm z_{k}}\pm c_k\Delta_k,
\label{POINTS}
\end{equation}
where $c_k$ is a positive gain coefficient and $\Delta_k$ is a vector whose components are randomly generated at each iteration from the set $\{1,-1,i,-i\}$. CSPSA allows for the existence of noise $\zeta_{k,\pm}$ in the evaluations $f({\bm z_{k,\pm}})$.

The gain coefficients are defined by the sequences
\begin{equation}
a_{k} = a/(k+A)^s
\end{equation}
and
\begin{equation}
c_k = b/k^r,    
\end{equation}
where $\{s, r, A, a, b\}$ are gain parameters. The values of the gain parameters are chosen to achieve the best possible rate of convergence. Therefore, the selection of the values itself becomes a costly optimization problem whose solution depends on the objective function and the particular optimizer. To avoid this problem, two sets of gain parameters are commonly used. Standard gain parameters with $s=0.602$, $r=0.101$, $A=10000.0$, $a=2.25$ and $b=0.5$, which provide fast convergence in the regime of a small number of iterations, and asymptotic gain parameters with $s=1.0$, $r=0.166$, $A=0.0$, $a=2.0$ and $b=0.5$, which provide fast convergence in the regime of a large number of iterations.

\section{Results}

We start to analyze our approach by considering the simplest case, namely, the discrimination of two unknown orthogonal pure states. We assume that Alice prepares the states $\{|0\rangle,|1\rangle\}$ with a priori probabilities $\eta_0$ and $\eta_1$, respectively. These orthogonal states are transformed by the communication channel into the states,
\begin{align}
    \ket{\psi_0} = \sqrt{\frac{1+s}{2}} \ket{0} + \sqrt{\frac{1-s}{2}} \ket{1}, \\
    \ket{\psi_1} = \sqrt{\frac{1+s}{2}} \ket{0} - \sqrt{\frac{1-s}{2}} \ket{1},
    \label{Two-pure-states}
\end{align}
where the parameter $s$ corresponds to the real-valued inner product $\langle\psi_0|\psi_1\rangle$. In this scenario, the optimal average error probability is given by the Helstrom bound \cite{Helstrom}
\begin{equation}
    p_{err} = \frac{1}{2} (1 - \sqrt{1 - 4\eta_0\eta_1s^2 }),
    \label{Helstrom-bound}
\end{equation}
which can be achieved by measuring an observable.

\begin{figure}
    \centering
    \begin{subfigure}[b]{\linewidth}
	    \includegraphics[width=\linewidth]{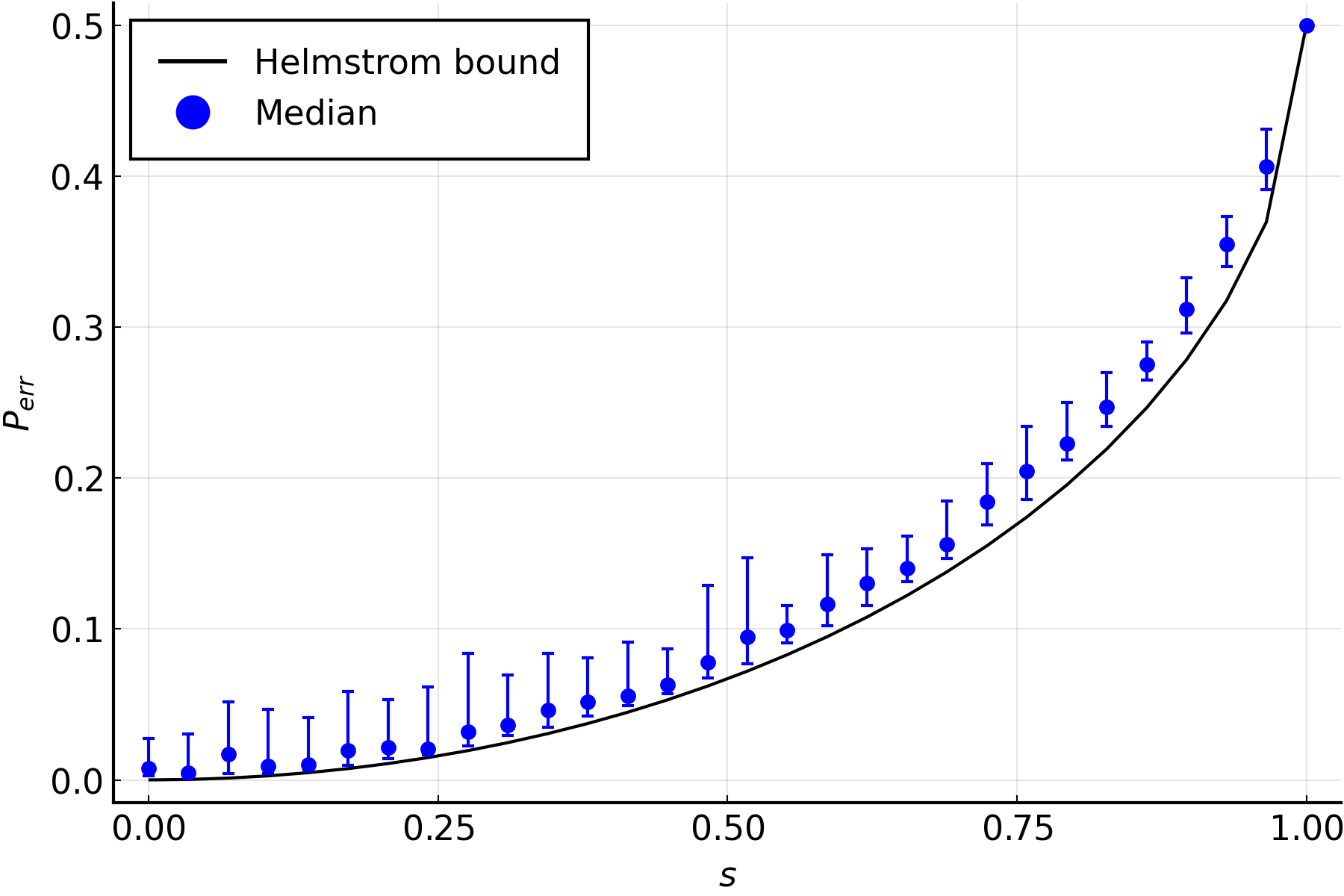}
	    \caption{}\label{fig:helmstrom_n150it100}
	\end{subfigure}
    \begin{subfigure}[b]{\linewidth}
    	\includegraphics[width=\linewidth]{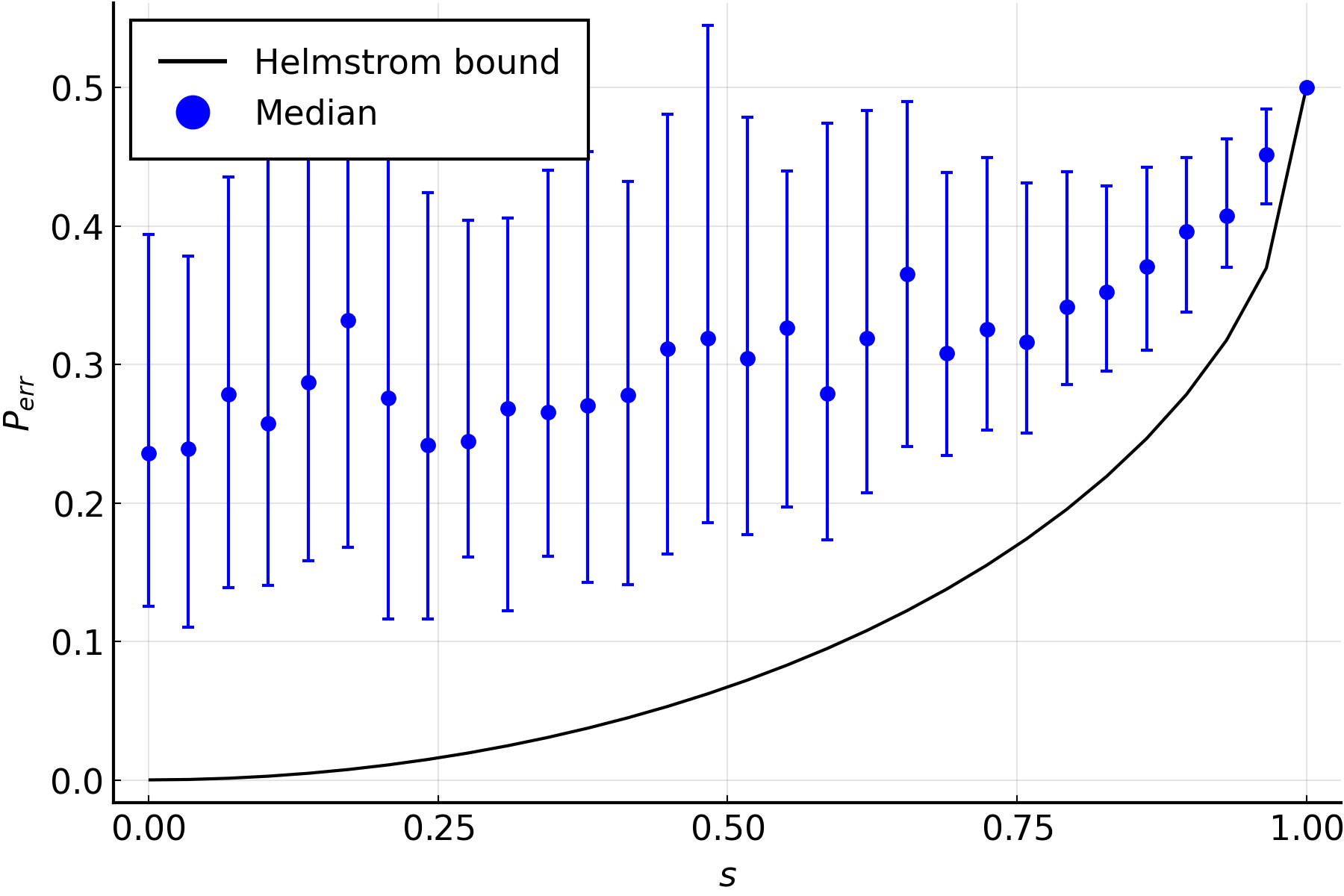}
    	\caption{}\label{fig:helmstrom_n1500it10}
    \end{subfigure}
    \begin{subfigure}[b]{\linewidth}
    	\includegraphics[width=\linewidth]{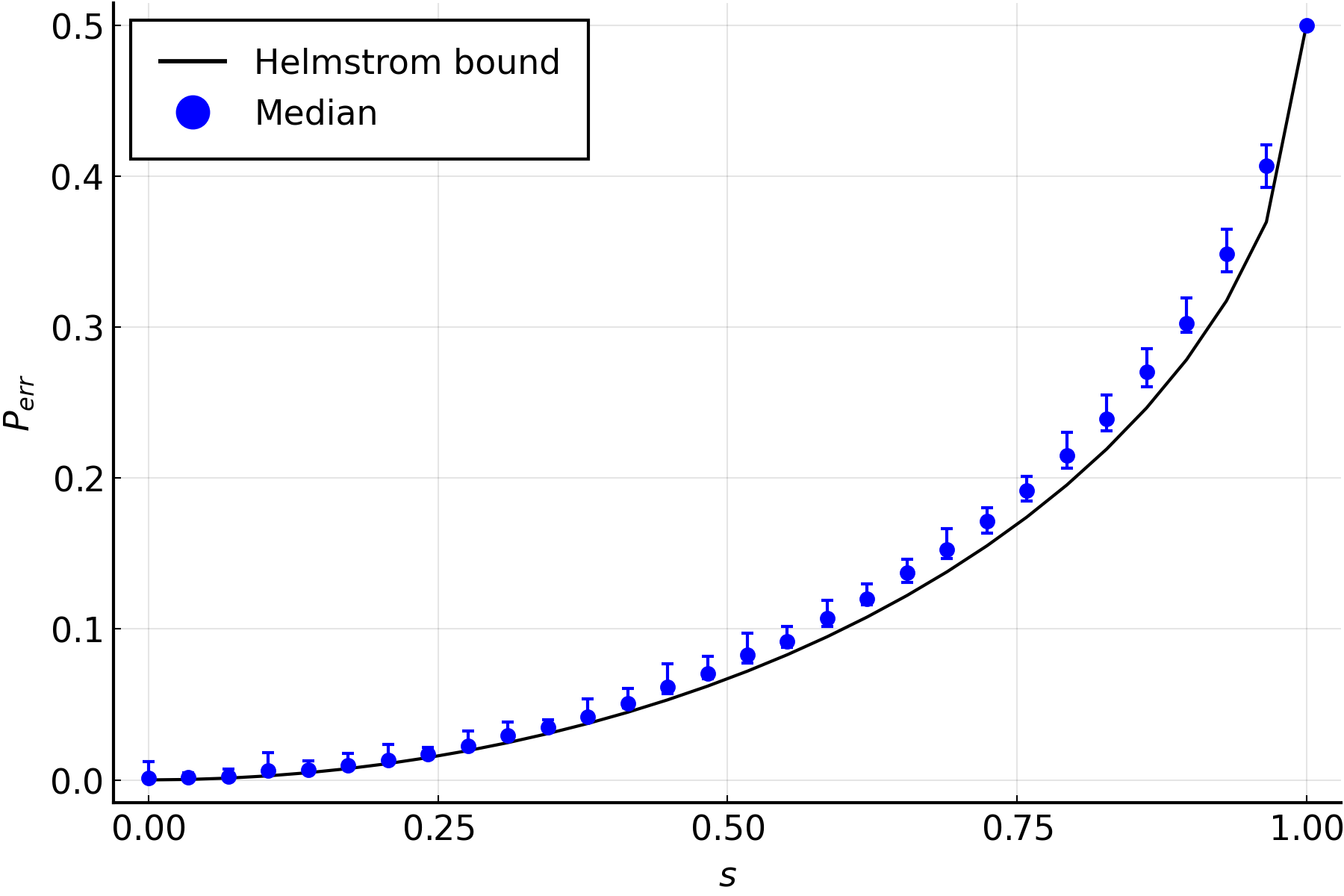}
    	\caption{}\label{fig:helmstrom_n50it300}
    \end{subfigure}
    \caption{Median of the estimated minimum-error probability $\tilde{p}_{err}$ as a function of the inner product $s$ between two unknown pure states given by Eq.~(\ref{Two-pure-states}) for $\eta_0=\eta_1=1/2$. Solid black line corresponds to the minimum-error probability $p_{err}$ given by the Helstrom bound in Eq.~(\ref{Helstrom-bound}). Solid blue dots corresponds to the median of $\tilde{p}_{err}$ calculated over 100 initial conditions for each value of $s$. Blue error bars indicate interquartile range.(a) $N=150$ and $k_t=100$, (b) $N=1500$ and $k_t=10$, and (c) $N=50$ and $k_t=300$. Asymptotic gain parameters are used.}
    
    \label{fig:helmstrom}
\end{figure}

We assume that the value of $s$ is unknown. The training of the measurement device is carried out without the use of a priori information. In particular, the training does not use the facts that the transmitted states are pure and that the optimal measurement is an observable. For a given value of $s$ our training procedure leads to a quantum measurement characterized by a value $\tilde{p}_{err}$ close to the optimal value given by the Helstrom bound. This is depicted in Fig.~\ref{fig:helmstrom_n150it100}, which shows the value of $\tilde{p}_{err}$ achieved by the training procedure as a function of $s$ for $\eta_0=\eta_1=1/2$. Since the optimization algorithm is stochastic, for each value of $s$ we repeat the procedure considering 100 randomly chosen initial guesses in the control parameter space and 100 iterations. The statistic generated by each POVM is simulated using an ensemble size $N=150$. In Fig.~\ref{fig:helmstrom_n150it100} the solid black line describes the Helstrom bound while the solid blue dots indicate the median of $\tilde{p}_{err}$ calculated over the 100 repetitions. The blue error bars describe the interquartile range. As is apparent from this figure, the training of the measurement device provides a median of $\tilde{p}_{err}$ that is very close to the Helstrom bound, where the difference $|\tilde{p}_{err}-p_{err}|$ is on the order of $10^{-2}$. The training procedure leads to similar results for other values of generation probabilities.

The total ensemble $N_t$ employed throughout the training process is given by $N_t=2N k_t$, where $k_t$ is the total number of iterations. Thus, the total ensemble can be split among the total number of iterations and the ensemble used to estimate the statistics of each POVM. This raises the question whether for a fixed total ensemble better accuracy is achieved by increasing $k_t$ or $N$. Figures~\ref{fig:helmstrom_n1500it10} and \ref{fig:helmstrom_n50it300} show the impact on $\tilde{p}_{err}$ for different splittings of $N_t=15\times10^3$. In Fig.~\ref{fig:helmstrom_n1500it10} we have $N=1500$ and $k_t=10$ and in Fig.~\ref{fig:helmstrom_n50it300} we have $N=50$ and $k_t=300$. As these two figures indicate, a much better accuracy is obtained by splitting the total ensemble in a small ensemble $N$ and a large number $k_t$ of iterations. In particular, in Fig.~\ref{fig:helmstrom_n50it300} the difference $|\tilde{p}_{err}-p_{err}|$ is in the order of $10^{-3}$, that is, one order of magnitude smaller than in the case of Fig.~\ref{fig:helmstrom_n150it100}. Furthermore, the interquartile range becomes narrower indicating less variability in the set of estimates $\{\tilde{p}_{err}\}$ for a given $p_{err}$.

\begin{figure}
	\centering
	\begin{subfigure}[b]{\linewidth}
		\includegraphics[width=\linewidth]{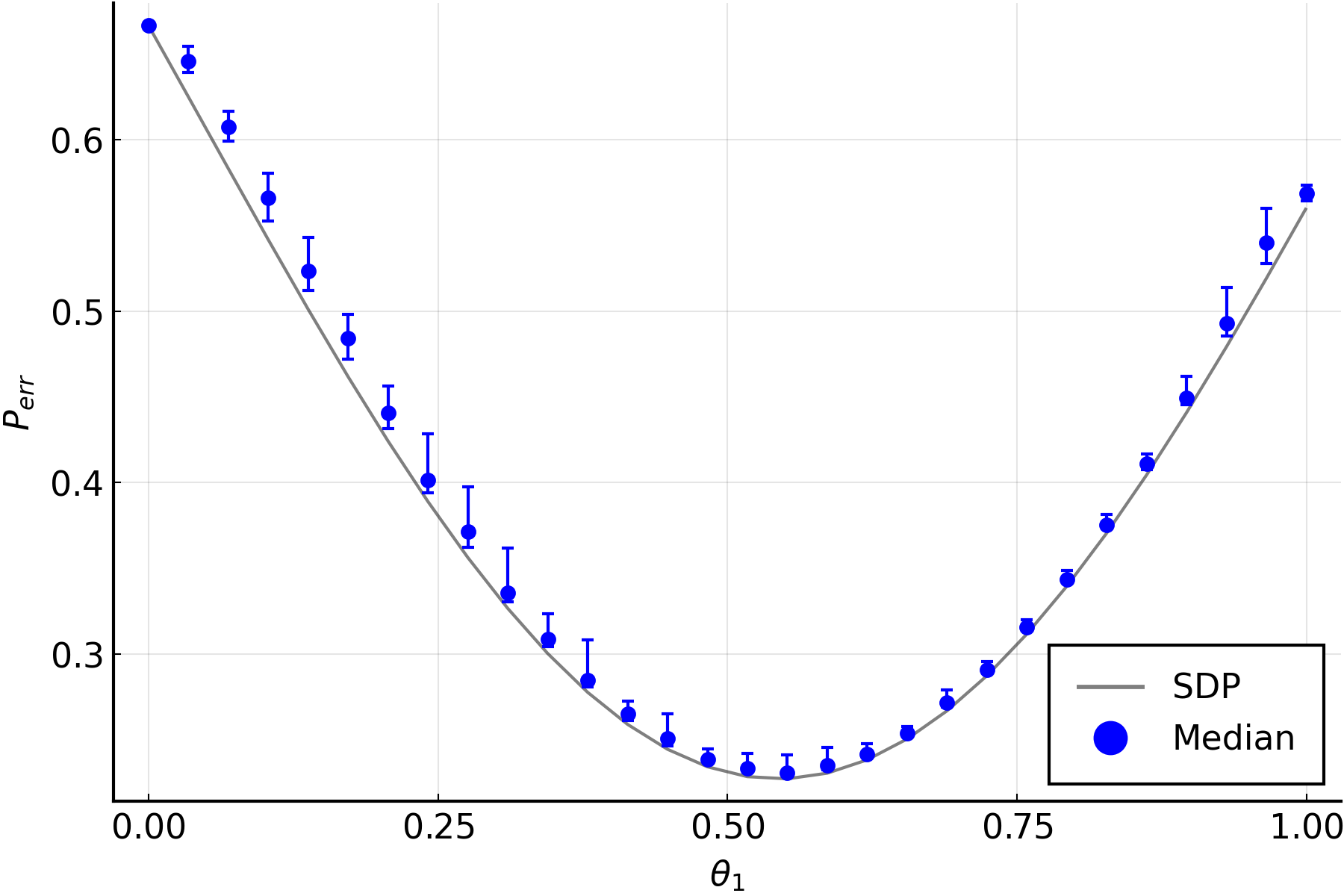}
		\caption{}\label{fig:n3d3}
	\end{subfigure}
	\begin{subfigure}[b]{\linewidth}
		\includegraphics[width=\linewidth]{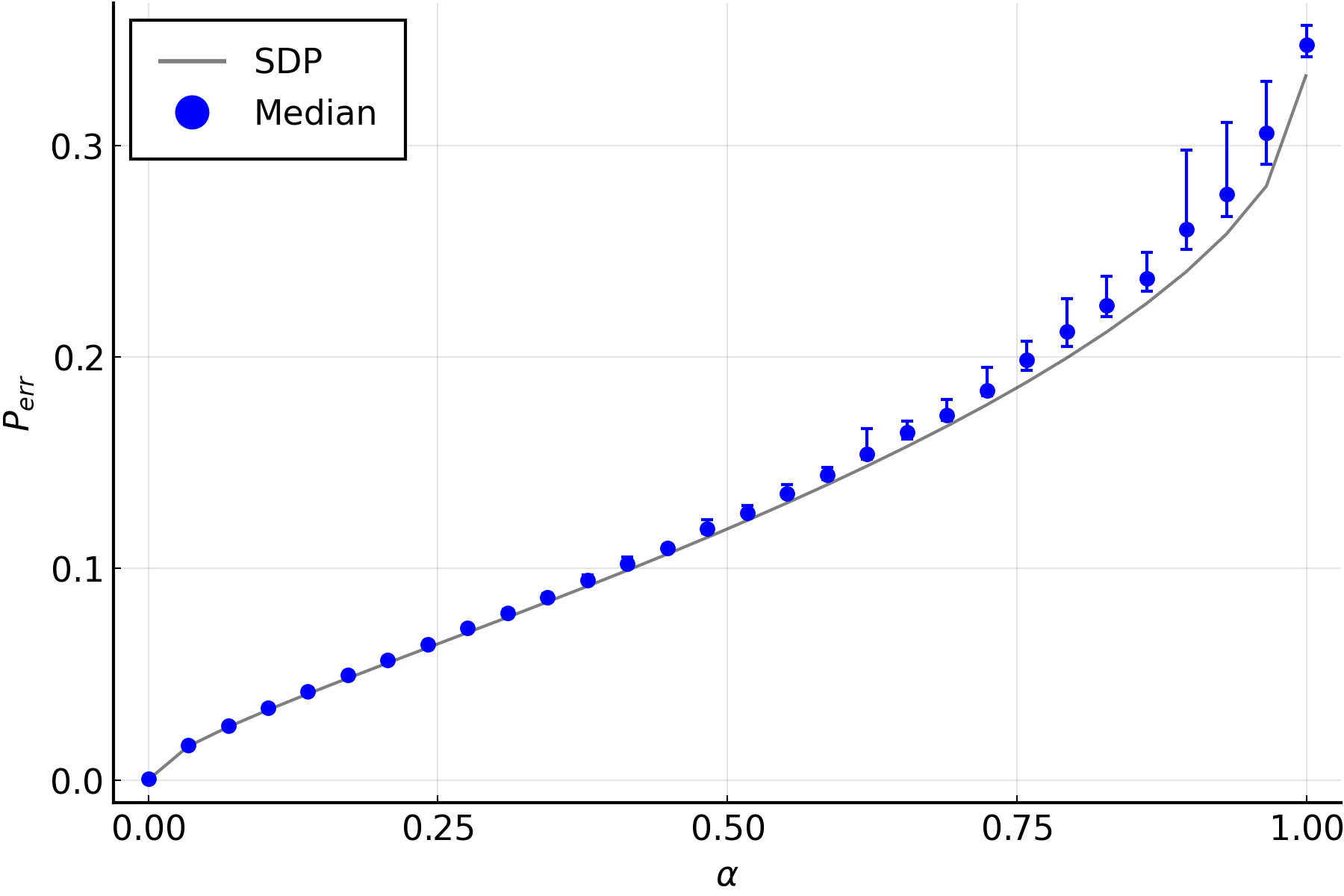}
		\caption{}\label{fig:n4d4} 
	\end{subfigure}
	\begin{subfigure}[b]{\linewidth}
		\includegraphics[width=\linewidth]{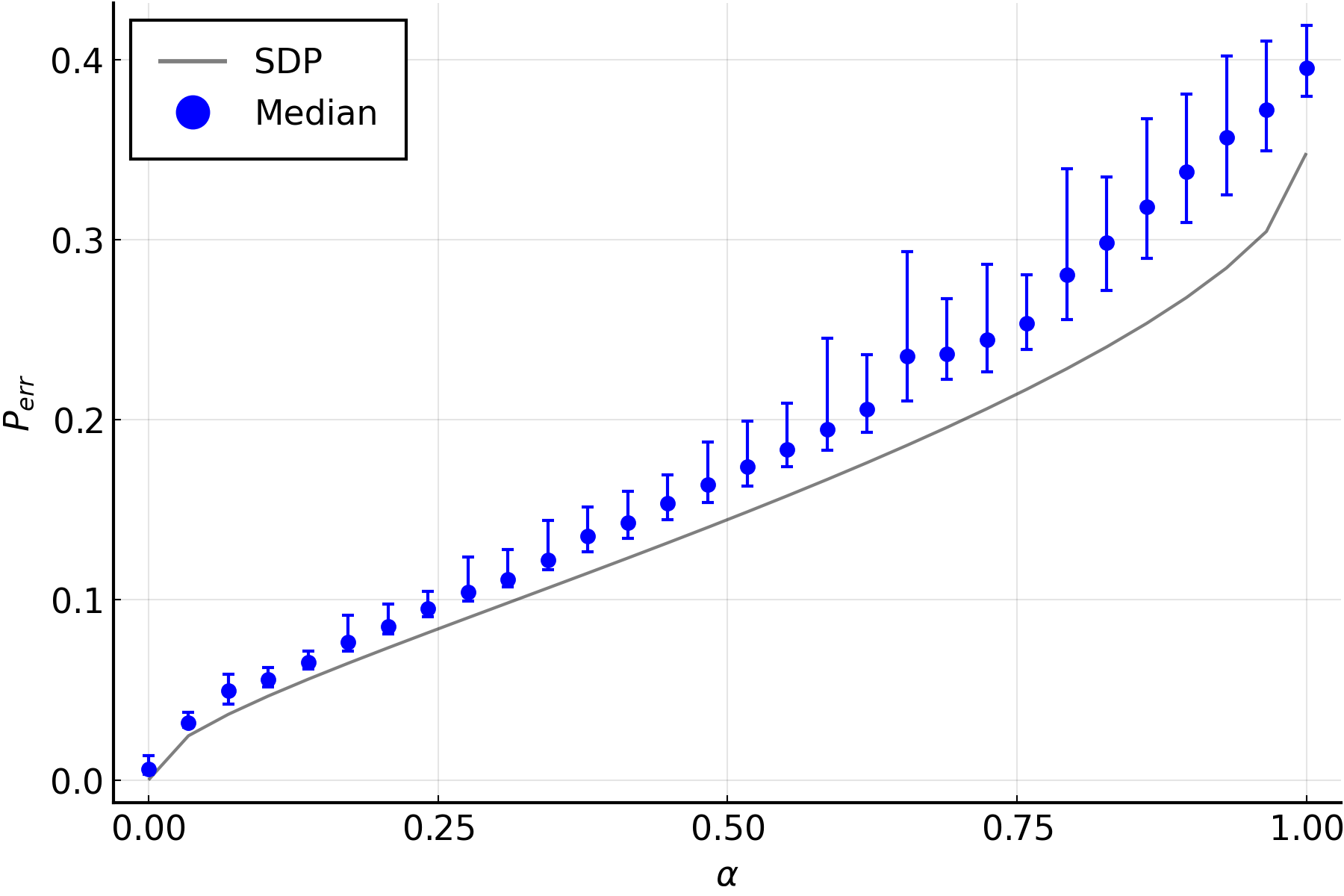}
		\caption{}\label{fig:n5d5} 
	\end{subfigure}
	
	\caption{Median of the estimated minimum-error probability $\tilde{p}_{err}$ for symmetric states as a function of: (a) $\theta_1$ for $d=3$, (b) $\alpha$ for $d=4$, and (c) $\alpha$ for $d=5$. Solid black line corresponds to the optimal minimum-error probability $p_{err}$. Solid blue dots correspond to the median of $\tilde{p}_{err}$ calculated over 100 initial conditions for each set of symmetric states. Blue error bars indicate interquartile range. (a) $N=300$ and $k_t=300$. (b) $N=300$ and $k_t=6\times10^3$. (c) $N=300$ and $k_t=1.2\times10^4$. Asymptotic gain parameters are used.}
	
	\label{fig:symmetric-states}
\end{figure}

The training of the measurement device for the discrimination of a larger number of states is demonstrated via symmetric states. These are given by the expression
\begin{equation}
    \ket{\psi_j} = \sum_{m=0}^{d-1} c_m \omega^{jm} \ket{m},~j \in \{ 0, \cdots, d-1 \},
    \label{d-Symmetric-states}
\end{equation}
where $d$ is the dimension of ${\cal H}_s$, $\omega = \exp{(2\pi i / d)}$, and the coefficients $c_m$ are constrained by the normalization condition. As long as the generation probabilities are equal, $d$ non-orthogonal symmetric states can be identified by measuring an observable whose eigenstates are given by the Fourier transform of the canonical base $\{|m\rangle\}$ (with $m=0,\dots,d-1$). The minimum-error discrimination of symmetric states has been experimentally demonstrated with high accuracy in dimensions up to $d=21$ \cite{Prosser}. The discrimination of symmetric states typically arises in the processes of quantum teleportation, entanglement swapping and dense coding. These use a maximally entangled quantum channel as resource. If the entanglement decreases along the generation of the channel, then the performance of the process can be enhanced by resorting to the local discrimination of symmetric states, where the coefficients $c_k$ entering in Eq.~(\ref{d-Symmetric-states}) are given by the real coefficients of the partially entangled state. If in addition, the channel coefficients are unknown, then our approach can be used. 

In the case of three symmetric states, the channel coefficients are parameterized as $c_0=\cos(\theta_1/2)\cos(\theta_2/2)$ and $c_1=\sin(\theta_1/2)\cos(\theta_2/2)$ with $\theta_1$ and $\theta_2$ in the interval $[0,\pi]$. Figure~\ref{fig:n3d3} shows $\tilde{p}_{err}$ as a function of $\theta_1$ for a particular value of $\theta_2$. The solid black line corresponds to the optimal minimum error discrimination probability $p_{err}$ while the solid blue dots indicate the median of $\tilde{p}_{err}$ calculated on 100 initial conditions for each value of $\theta_1$ after $10^3$ iterations using an ensemble size $N=10^3$. The difference $|p_{err}-\tilde{p}_{err}|$ is on the order of 
$10^{-2}$, as in the case of Fig.~\ref{fig:helmstrom}, but is obtained with a higher number of iterations and a larger ensemble size. In the case of a higher number of states we resort to a bi-parametric family of symmetric states given by $c^2_k\propto 1-\sqrt[d]{[(k-j_0+1)/(d-j_0)]\alpha}$ if $k\ge j_0$ and $c^2_k\propto 1$ if $k<j_0$, where $j_0=1,\dots,d-1$ and $\alpha\in[0,1]$. Figures~\ref{fig:n4d4} and \ref{fig:n5d5} show the behavior of $\tilde{p}_{err}$ as a function of $\alpha$ and $j_0=2$ for $d=4$ and $d=5$, respectively. In both cases the ensemble size is $N=300$ and the median was calculated over 100 initial conditions for each value of $\alpha$. As in Fig.~\ref{fig:n3d3}, the difference $|p_{err}-\tilde{p}_{err}|$ is on the order of $10^{-2}$. To achieve this result, however, it was necessary to increase the number of iterations to $6\times10^3$ and $1.2\times10^4$ for $d=4$ and $d=5$, respectively. 

Thus, as we increase the number $n$ of states to be discriminated and the dimension $d$ of the Hilbert space, the number of iterations $k_t$ and the ensemble size $N$ required to achieve a given tolerance also increase. This is due to the fact that the dimension of the search space, that is, the number of parameters that control the measurement device, increases as $nd^2$. In addition, the probabilities entering in the estimate of $p_{err}$ of Eq.~(\ref{Estimate-minimum-error}) are obtained using as a resource a given ensemble size. As the number of probabilities increases it is necessary to increase the ensemble size $N$ to obtain probability estimates that lead to a given tolerance.

So far we have considered training the measurement device assuming the most general quantum measurement. This typically conveys an increase in the resources required for the training. As our previous simulations indicate, as we increase the number $n$ of states to discriminate, as well as the dimension $d$, the training of the measurement device consumes even more resources, that is, larger ensembles and higher iteration numbers. To reduce the resources required for training, it is customary to reduce the dimension of the search space. This is done by imposing a set of conditions on the measurement device. This occurs when we have a priori information that allows us to ascertain that the optimal measurement satisfy a given condition. For instance, if the non-orthogonal states to be discriminated via the minimum-error strategy are pure and $n=d$, then we can assume that the optimal POVM is an observable. This effectively decreases the dimension of the search space. Another possibility is that we are interested in reaching a given value of the minimum-error probability in a particular family of measurement devices, in which case we don't need the optimal measurement.

\begin{figure}
    \centering
    \begin{subfigure}[b]{\linewidth}
    	\includegraphics[width=\linewidth]{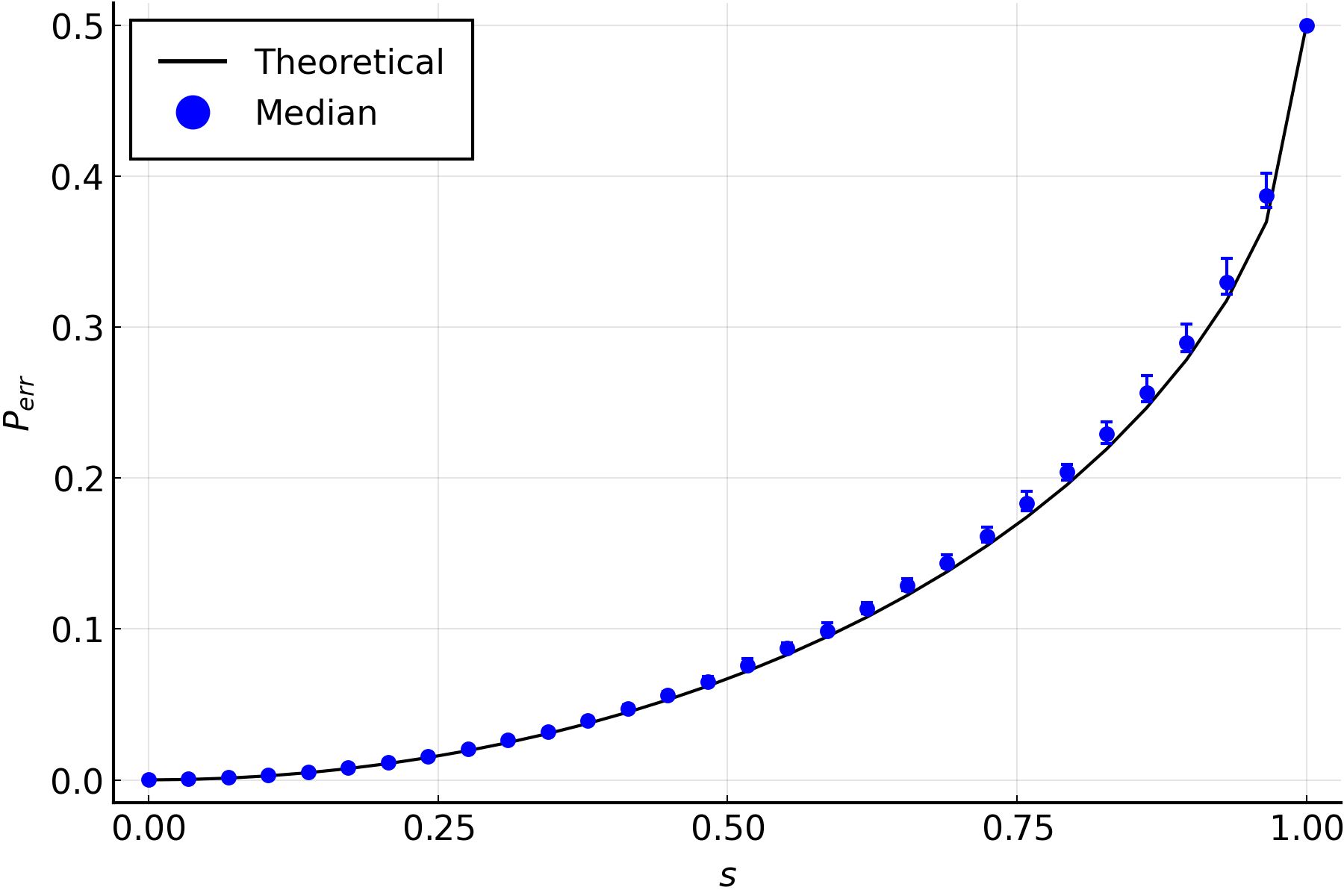}
    	\caption{}\label{fig:n2d2} 
    \end{subfigure}
	\begin{subfigure}[b]{\linewidth} 	
		\includegraphics[width=\linewidth]{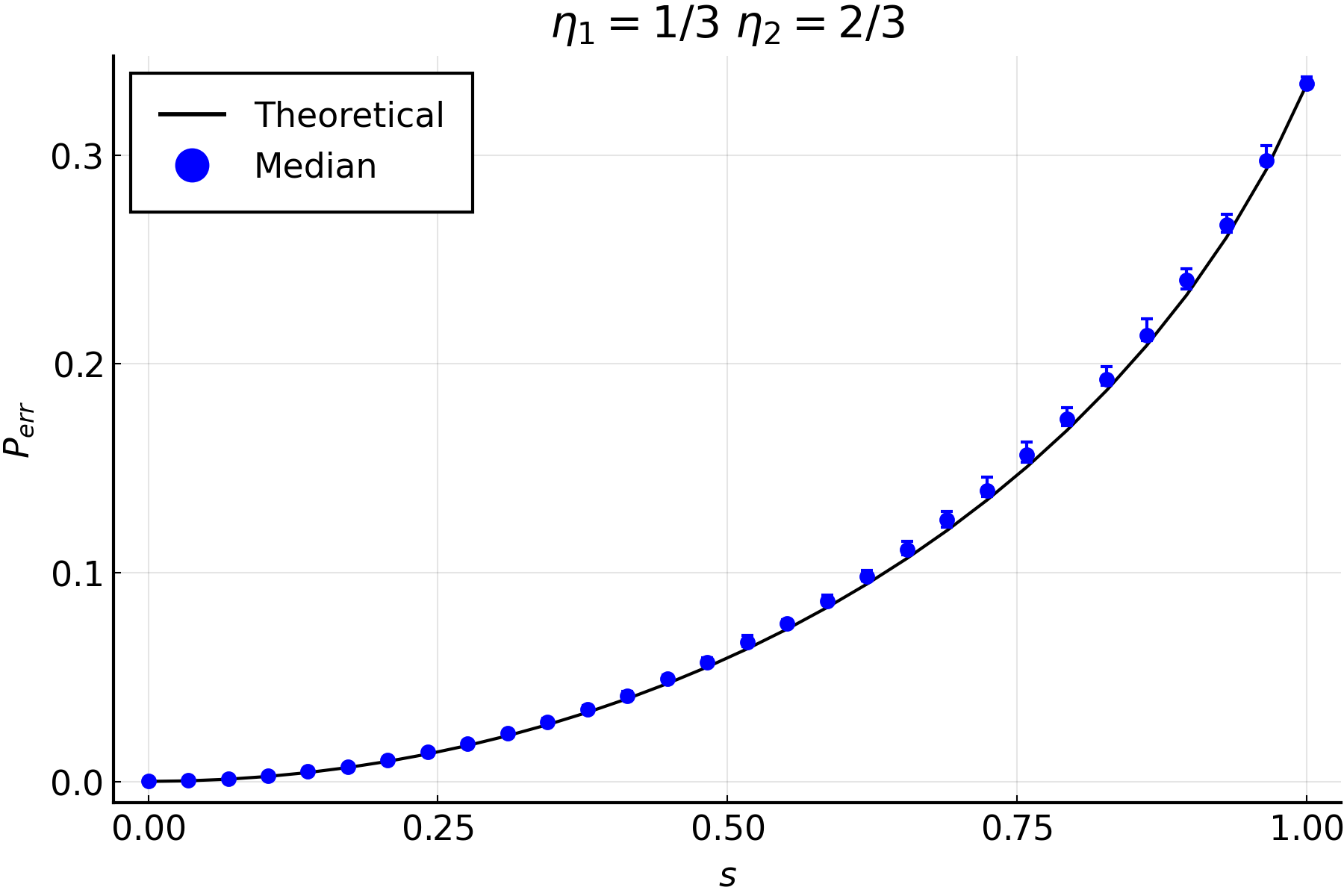}
		\caption{}\label{fig:n2d2-1}
	\end{subfigure}
    \begin{subfigure}[b]{\linewidth}
    	\includegraphics[width=\linewidth]{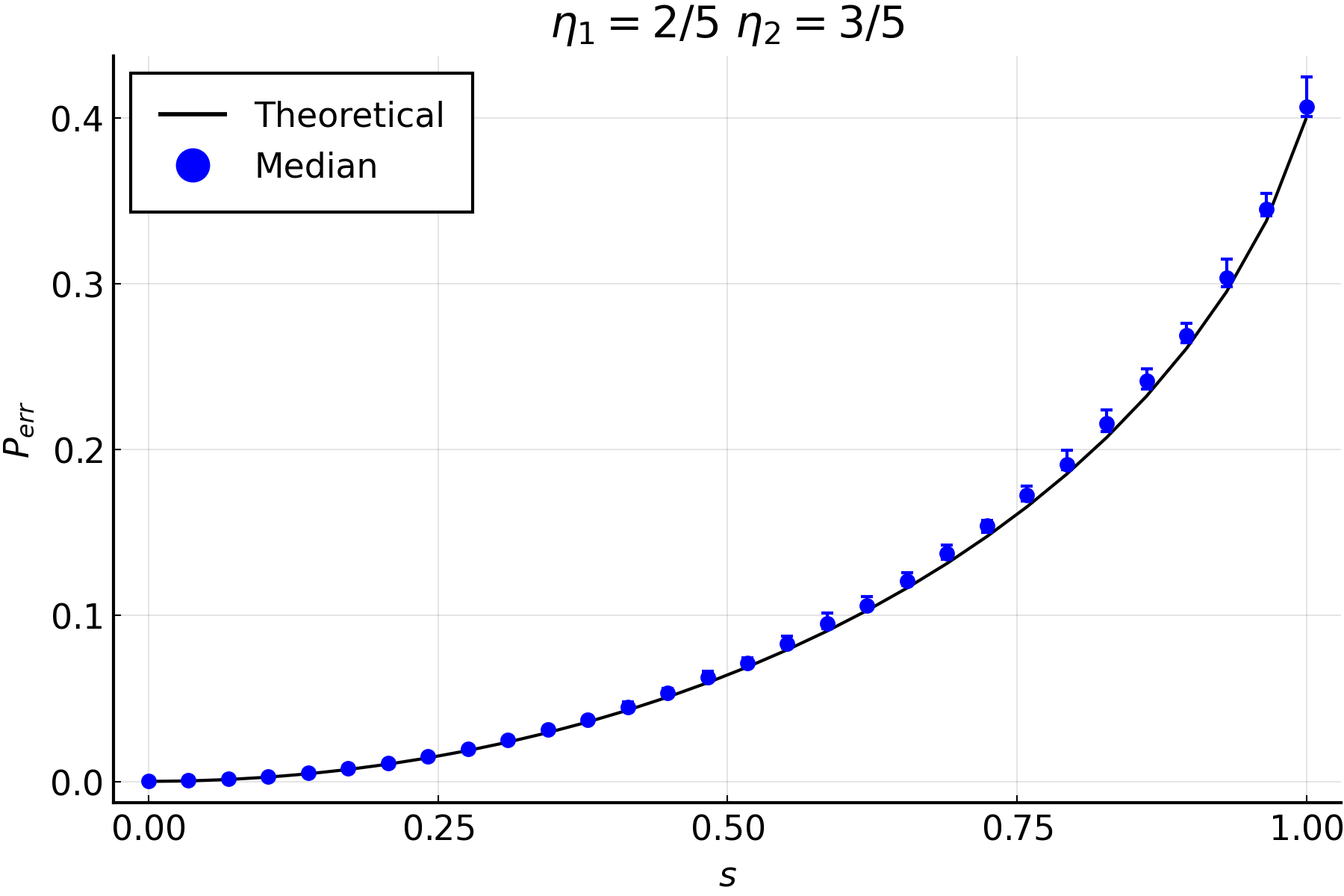}
    	\caption{}\label{fig:n2d2-2}
    \end{subfigure}
    \caption{Median of the estimated minimum-error probability $\tilde{p}_{err}$ as a function of the inner product $s$ between two unknown pure states given by Eq.~(\ref{Two-pure-states}). Solid black line corresponds to the minimum-error probability $p_{err}$ given by the Helstrom bound of Eq.~(\ref{Helstrom-bound}). Solid blue dots indicate the median of $\tilde{p}_{err}$ calculated over 100 initial conditions for each value of $s$. Blue error bars indicate interquartile range. (a) $\eta_0=\eta_1=1/2$, (b) $\eta_0=1/3$ and $\eta_1=2/3$, and (c) $\eta_0=2/5$ and $\eta_1=3/5$. Simulations are carried out with an ensemble size $N=50$ and total number of iterations $k_t=50$. Asymptotic gain parameters are used.}
    
    \label{fig:helstrom-observable}
\end{figure}

This is depicted in Fig.~\ref{fig:helstrom-observable}, where we reproduce the Helstrom bound for states in Eq.~(\ref{Two-pure-states}) by optimizing in the set of observables. In Fig.~\ref{fig:n2d2} we show the case of equal generation probabilities. In this case, the training was carried out using an ensemble size $N=50$ and a total number of iterations $k_{t}=50$, which leads to a difference $|p_{err}-\tilde{p}_{err}|$ is on the order of $10^{-3}$. This result can be compared to the one illustrated in Fig.\ref{fig:helmstrom_n50it300}, where the same ensemble size is used but with a much larger number of iterations $k_t=300$. Therefore, the reduction in the dimension of the search space leads to a reduction of the total ensemble $N_{tot}$ by a factor 6. A similar result holds in Figs.~\ref{fig:n2d2-1} and \ref{fig:n2d2-2} for different values of the generation probabilities. Let us note that the initial condition in the search space is randomly chosen, that is, we do not assume as initial condition an observable close to the optimal one.

\begin{figure}
    \centering
    \begin{subfigure}[b]{\linewidth}
    	\includegraphics[width=\linewidth]{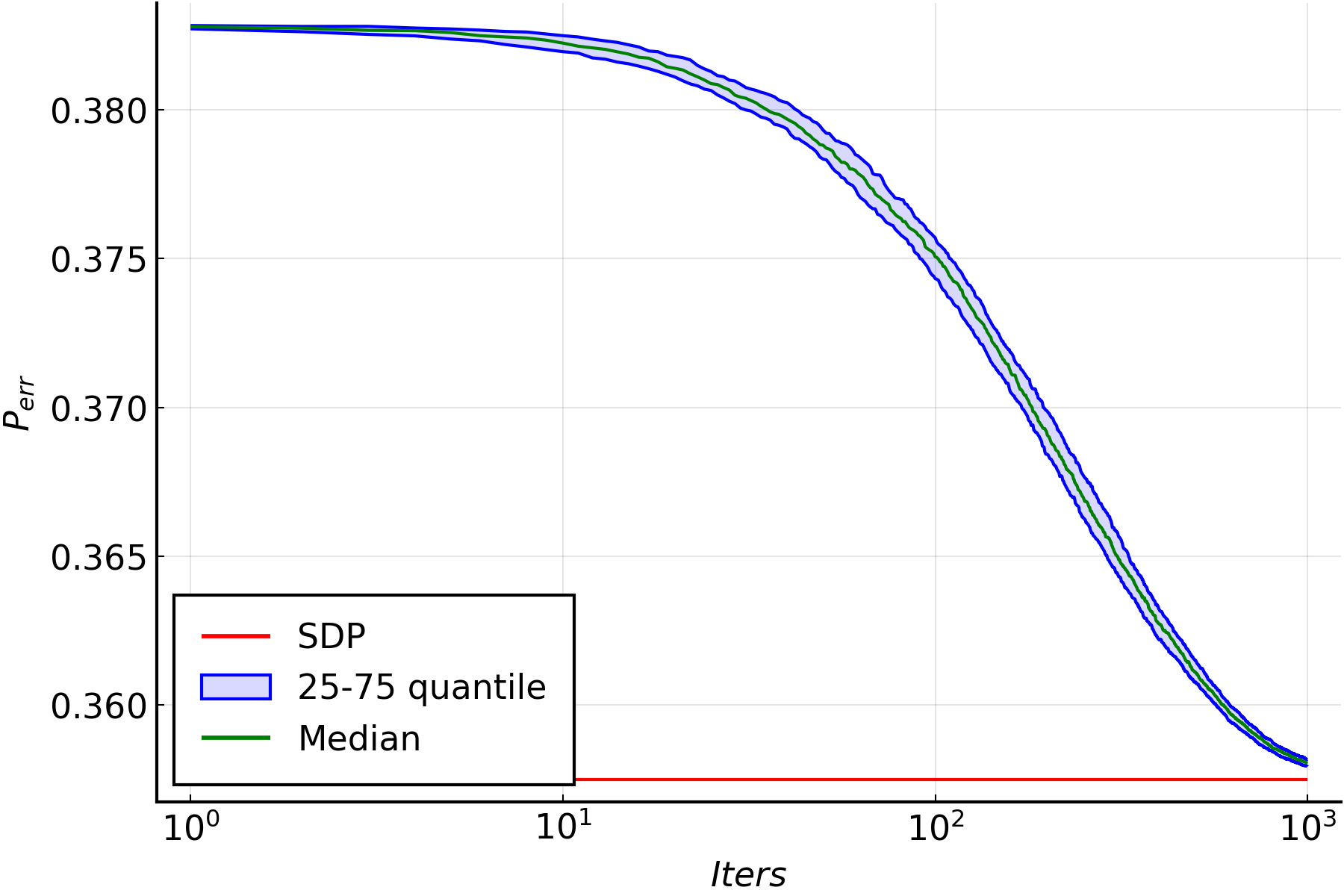}
    	\caption{} \label{fig:4-1}
    \end{subfigure}
    \begin{subfigure}[b]{\linewidth}
    	\includegraphics[width=\linewidth]{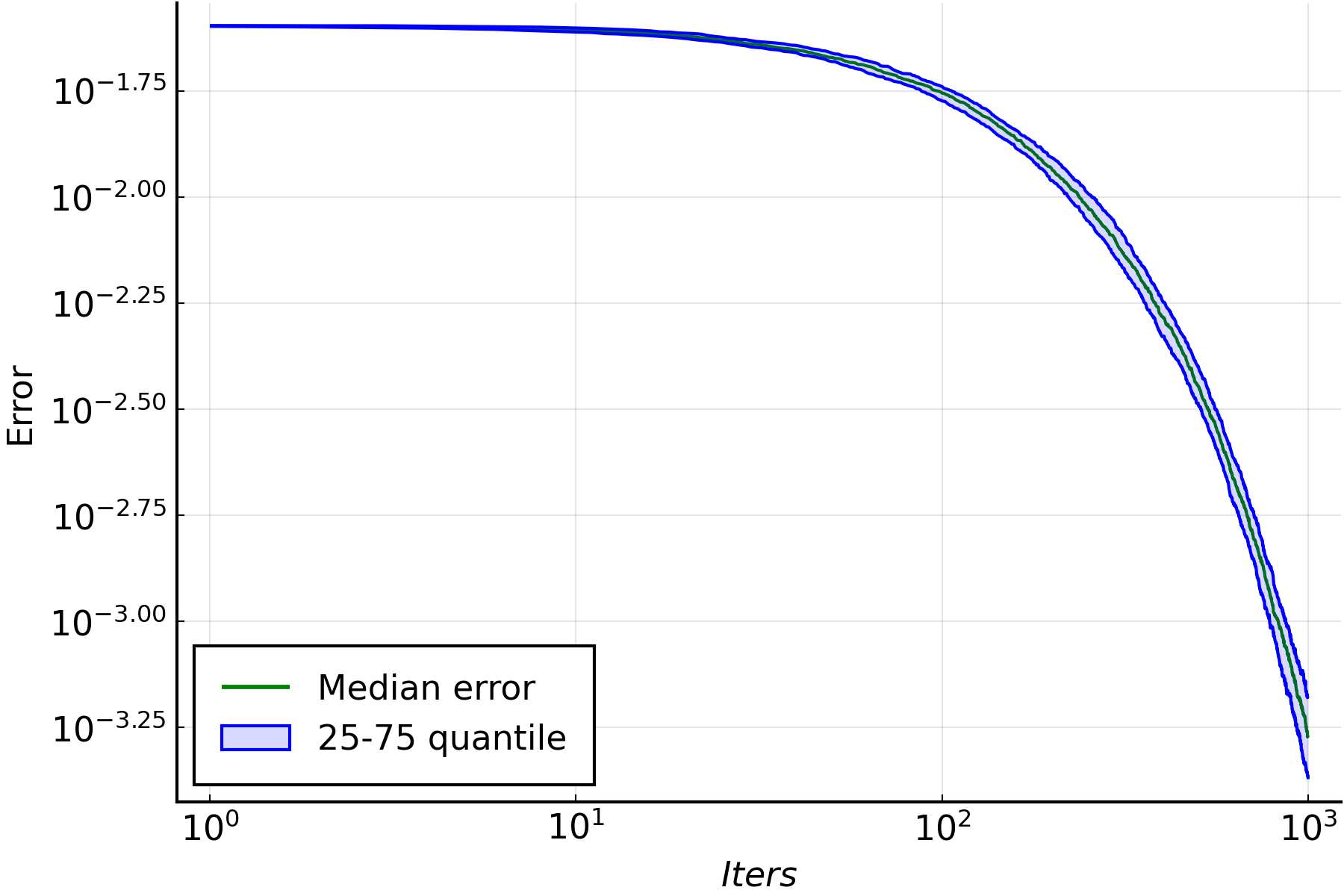}
    	\caption{} \label{fig:4-2}
    \end{subfigure}
    \caption{Discrimination of two unknown non-orthogonal mixed single-qubit states generated by a phase flip channel. (a) Solid green line indicates median value of $\tilde{p}_{err}$, calculated over 100 repetitions, as a function of the number of iterations. Solid red line corresponds to the optimal value of $p_{err}$ obtained via semidefinite programming. (b) Solid green line indicates the median value of $|\tilde{p}_{err}-p_{err}|$ calculated over 100 repetitions, as a function of the number of iterations. Shaded blue area corresponds to interquartile range and ensemble size $N=300$. Standard gain parameters are used.}
    \label{fig:qubitdephase}
\end{figure}

Finally, we consider a more realistic scenario in which two single-qubit orthogonal states emitted by Alice are subjected to the action of a phase flip channel \cite{Nielsen}. The action of this channel onto a single-qubit density matrix $\rho$ is defined by the relation
\begin{equation}
    \varepsilon (\rho) = p \rho + \left( 1-p \right) \sigma_z \rho \sigma_z,
\end{equation}
where the parameter $p$ represent the strength of the channel and $\sigma_z=|0\rangle\langle0|-|1\rangle\langle1|$. The phase flip channel nullifies the off-diagonal terms in the density operator with respect to the canonical basis $\{|0\rangle,|1\rangle\}$ while decreasing the purity. We assume that the states transmitted by Alice are random pure states and that the value of $p=3/5$ is unknown. The fidelity between the pure states and the noisy states is 0.785.

Figure \ref{fig:qubitdephase} shows the result of training the measurement device to discriminate the states generated by the phase flip channel. The training was initialized with the measurement that optimally discriminates the pure states sent by Alice. In Fig.~\ref{fig:4-1} displays the median (continuous green line) of $\tilde{p}_{err}$, calculated over 100 independent repetitions, as a function of the number of iterations. The continuous red line corresponds to the solution $p_{err}$ of the optimization of the minimum error for the states generated by the phase flip channel via semidefinite programming. Clearly, the value of $\tilde{p}_{err}$ obtained by training the measurement device converges to the optimal value $p_{err}$. The interquartile range, described by the shaded area, is very narrow indicating a very small variability of the training with respect to the initial conditions. Similar results hold for other values of the parameter $p$, which controls the convergence rate toward the optimal value of $p_{err}$. Figure~\ref{fig:4-2} displays the median of  $|\tilde{p}_{err}-p_{err}|$ as a function of the number of iterations.

\section{Conclusions}

Here, we have studied the problem of discriminating unknown non-orthogonal quantum states. This situation occurs when two parties try to transmit information encoded in orthogonal quantum states that are transformed into non-orthogonal states by the action of a partially characterized quantum channel. Since the communicating parties do not know the states generated by the channel, standard approaches to discriminate non-orthogonal quantum states cannot be applied. Instead, we have proposed to train a single-shot measurement to optimally discriminate unknown non-orthogonal quantum states. This device is controlled by a large set of parameters, such that a given set of parameter values corresponds to the realization of a positive operator-valued measure (POVM). The measurement device is iteratively optimized in the space of the control parameters, or search space, to achieve the minimum value of the error probability, that is, we seek to implement the minimum-error discrimination strategy. The optimization is driven by a gradient-free stochastic optimization algorithm that approximates the gradient of the error probability by a finite difference. This requires at each iteration evaluation of the error probability at two different points in the search space. Thereby, the training is driven by experimentally acquired data. The choice of a stochastic optimization methods is based on its robustness against noise.

We have studied the proposed approach using numerical simulations. First, we have shown that our approach leads to values of the error probability that are very close to the optimum. This was done in the case of two 2-dimensional unknown non-orthogonal pure states, where the optimal value of the average error probability is given by the Helstrom bound. Since the training method requires the estimation of probabilities, the total ensemble is regarded as a resource. This is divided evenly throughout the iterations of the training method. We have shown that best results, that is, a value of the error probability closer to the Helstrom bound, can be obtained for a fixed total ensemble size by increasing the number of iterations. Thereafter, we have extended our result to the case of $d$ $d$-dimensional symmetric states for $d=3,4,5$, where our method also provides accurate results. However, to achieve a fixed accuracy as we increase the number of states and the dimension, its necessary to increase the ensemble size and, consequently, the number of iterations. To avoid this, we have reduced the dimension of the search space by assuming that the required measurement has some special property. In particular, we have assumed that the optimal measurement is an observable. Thereby, in the case of two non-orthogonal pure states we have achieved a considerable reduction by a factor $1/6$ in the ensemble size, which leads to an equal reduction in the number of iterations. Finally, we have applied the training procedure to the phase flip channel and shown that it is possible to achieve a value of the error probability close to the optimal one. Note that our proposal does not require data post-processing methods, such as maximum likelihood or Bayesian inference, which helps reduce computational cost and avoids exponential scaling of multipartite quantum states.

Our proposal finds applications whenever two parties intend to communicate through a channel whose characterization is difficult or costly. For instance, processes such as quantum teleportation, entanglement swapping, and dense coding, when performed through a partially entangled channel, can become a problem of local discrimination of non-orthogonal states \cite{PhysRevA.71.012303,PhysRevA.89.012337,PhysRevA.85.062322,PhysRevA.68.022310,PhysLettA.239,PhysRevA.87.052327}. If the description of the entangled channel is not available, then the states to be discriminated are unknown, in which case our method can also be applied. Recently, the problem of optimally discriminating between different configurations of a complex scattering system has been studied \cite{Bouchet} from the point of view of quantum state discrimination, where several non-orthogonal quantum states of light are associated to different hypotheses about an scattering system. These must be resolved with the best possible accuracy, which is limited by the Helstrom bound in the simplest case. Our training method can also be applied to this problem by finding the best average error probability.

\begin{acknowledgments}
This work was supported by ANID -- Millennium Science Initiative Program -- ICN17$_-$012. AD was supported by FONDECYT Grant 1180558. LP was supported by ANID-PFCHA/DOCTORADO-BECAS-CHILE/2019-72200275. LZ was supported by ANID-PFCHA/DOCTORADO-NACIONAL/2018-21181021.
\end{acknowledgments}

\bibliographystyle{apsrev4-2}
\bibliography{bibtex}

\end{document}